\documentclass{article}
\usepackage{graphics}
\usepackage{epsfig}
\begin{document}

\begin{center}
{\large \bf Single-photon emission via Raman scattering from the
levels with partially resolved hyperfine structure}
\end{center}
\bigskip
\begin{center}
\textbf{V.A. Reshetov}\\
\bigskip
\textit{Department of General and Theoretical Physics, Tolyatti
State University, 14 Belorousskaya Street, 446667 Tolyatti, Russia}
\end{center}
\begin{center}
\textbf{I.V. Yevseyev}\\
\bigskip
\textit{Department  of Theoretical Physics,  Moscow Engineering
Physics Institue, 31 Kashirskoe Shosse, 115409 Moscow, Russia}
\end{center}

\begin{abstract}
The probability of emission of a single photon via Raman scattering
of laser pulse on the three-level $\Lambda$ - type atom in
microcavity is studied. The duration of the pulse is considered to
be short enough, so that the hyperfine structure of the upper level
remains totally unresolved, while that of the lower level is totally
resolved. The coherent laser pulse is assumed to be in resonance
with the transition between one hyperfine structure component of the
lower atomic level and all hyperfine structure components of the
upper level, while the quantized cavity field is assumed to be in
resonance with the transition between the other hyperfine structure
component of the lower level and all components of the upper one.
The dependence of the photon emission probability on the mutual
orientation of polarization vectors of the cavity mode and of the
coherent laser pulse is analyzed. Particularly, the case is
investigated, when the total electronic angular momentum of the
lower atomic level equals 1/2, which is true for the ground states
of alkali atoms employed in the experiments on deterministic single
photon emission. It is shown, that in this case the probability of
photon emission equals zero for collinear polarizations of the
photon and of the laser pulse, and the probability obtains its
maximum value, when the angle between their polarizations equals
$60^{\circ}$.
\end{abstract}

\section{Introduction}

The implementation of quantum optical devices for quantum
information processing is a rapidly developing field of research
nowadays, passing the threshold from a learning phase into a domain
of rudimentary functionality \cite{n1}. Among such devices the
three-level $\Lambda$ - type atom placed inside a high-finesse
cavity proved to be a useful building block for quantum
communication schemes because of its ability to interface
efficiently atoms and photons \cite{n2,n3,n4}. As it was pointed out
in \cite{n5} and then realized experimentally in \cite{n6,n7,n8},
such atoms may be employed for the controlled generation of a single
photon in the cavity by means of vacuum-stimulated Raman scattering.
In these experiments one branch of the atomic $\Lambda$ - type
transitions was coupled to the quantized cavity field, while the
other one was coupled to the coherent laser field. In course of
resonant interaction with the fields inside the cavity the atom
emitted a single photon. The effectiveness of this process depends
on the interaction parameters, in particular, it depends on the
relative orientation of polarizations of the cavity field and of the
laser field, since the atomic levels are usually strongly degenerate
and the contributions of various Zeeman components of resonant
levels to the interaction are determined by the polarizations of the
fields. Thus in the experiment \cite{n3,n4} the transitions between
hyperfine components $F_{a}=3$ and $F'_{a}=4$ of the ground state
$6S_{1/2}$ and the component $F_{b}=4$ of the excited state
$6P_{3/2}$ of Cesium atoms were employed, while in the experiments
\cite{n6,n7,n8} they were the transitions between hyperfine
components $F_{a}=2$ and $F'_{a}=3$ of the ground state $5S_{1/2}$
and the component $F_{b}=3$ of the excited state $5P_{3/2}$ of
Rubidium atoms. Such polarization dependencies were studied
previously \cite{n9} for rather long interaction times, when the
hyperfine structure of both the ground and the excited states was
totally resolved. In this paper we study the polarization
dependencies of the probability of single-photon emission via Raman
scattering of rather short coherent laser pulses, such that the
hyperfine structure of the excited level appears to be totally
unresolved. However we consider the hyperfine structure of the
ground state to remain totally resolved, since the frequencies of
hyperfine splitting of ground states of alkali atoms are
significantly greater, than those of the excited states.  In section
2 we describe the interaction model and obtain the evolution
operator for this model, while in section 3 we obtain the expression
for the probability of single photon emission and analyze it
numerically for the transitions in Cesium and Rubidium, employed in
the experiments
 \cite{n3,n4} and  \cite{n6,n7,n8}.

\begin{figure}[t]
\includegraphics[width=7cm]{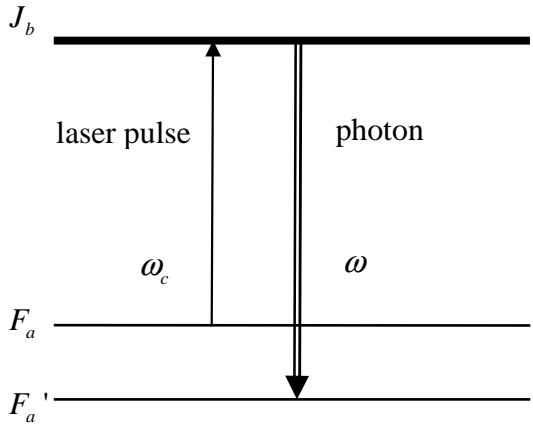}
\caption{The level diagram.}
\end{figure}

\section{Evolution operator}

We assume, that the coherent laser pulse with the carrier frequency
$\omega_{c}$ is in resonance with the transition $F_{a}\rightarrow
J_{b}$ between one hyperfine structure component $F_{a}$ of the
ground state $a$ and all the hyperfine structure components $F_{b}$
of the excited state $b$, while the quantized cavity field with the
carrier frequency $\omega$ is in resonance with the transition
$F'_{a}\rightarrow J_{b}$ between some other hyperfine structure
component $F'_{a}$ of the ground state and all the components of the
excited state (Fig.1). Here $F_{a}$ and $F_{b}$ are the values of
the atomic total angular momenta of the ground and excited states
respectively, while $J_{a}$ and $J_{b}$ denote electronic total
angular momenta of these states,
$F_{a,b}=|J_{a,b}-I|,...,J_{a,b}+I$, $I$ being the value of nuclear
spin. The electric field strength, which contains the coherent field
of the pulse and the quantized field of the cavity, may be put down
as:
     \begin{equation}\label{q1}
  {\bf \hat{E}\/}=e_{c}{\bf l\/}_{c} e^{-\imath \omega_{c} t}+
  \imath e_{0}{\bf l\/} \hat{a} e^{-\imath \omega t} + h.c.~,
     \end{equation}
where $e_{c}$ and ${\bf l\/}_{c}$ are the constant amplitude and the
unit polarization vector of the pulse, $\hat{a}$ is the photon
annihilation operator for the cavity mode, while
     \begin{equation}\label{q2}
  e_{0}= \sqrt{\frac{2\pi \hbar \omega}{V_{c}}},
     \end{equation}
is the photon field, $V_{c}$ being the cavity (quantization) volume,
${\bf l\/}$ - the unit polarization vector of the cavity field. The
quantum system consists of the atom and of the quantized field. The
equation for the slowly-varying density matrix $\hat{\rho}$ of this
system in the rotating-wave approximation is as follows:
     \begin{equation}\label{q3}
  \dot{\hat{\rho}}=\imath\left[\hat{V},\hat{\rho}\right],~
  \hat{V}=\chi_{c}\hat{p}_{c} - \imath \chi\hat{p}\hat{a}^{+} + h.c.
    \end{equation}
Here $\chi_{c}=|d|e_{c}/\hbar$ and $\chi=|d|e_{0}/\hbar$ are the
reduced Rabi frequencies for the coherent pulse and for the cavity
field, $d=d(J_{a}J_{b})$ being the reduced matrix element of the
electric dipole moment operator for the electronic transition
$J_{a}\rightarrow J_{b}$, while
    \begin{equation}\label{q4}
  \hat{p}_{c}=\hat{\textbf{g}}_{c}\textbf{l}_{c}^{*},~
  \hat{p}=\hat{\textbf{g}}\textbf{l}^{*},
    \end{equation}
$\hat{\textbf{g}}_{c}$ and $\hat{\textbf{g}}$ are the dimensionless
electric dipole moment operators for the transitions
$F_{a}\rightarrow J_{b}$ and $F'_{a}\rightarrow J_{b}$. The circular
components of these vector operators are expressed through Wigner
3J- and 6J- symbols and partial operators
    \begin{equation}\label{q5}
  \hat{P}^{F_{\alpha}F'_{\beta}}_{M_{\alpha}M'_{\beta}} =
  |F_{\alpha}M_{\alpha}><F'_{\beta}M'_{\beta}|,~ \alpha, \beta =a,b,
    \end{equation}
in a following way:
    \begin{equation}\label{q6}
  \hat{g}_{cq}=\sum_{M_{a},M_{b},F_{b}}
  (g_{cq})^{F_{a}F_{b}}_{M_{a}M_{b}}\cdot \hat{P}^{F_{a}F_{b}}_{M_{a}M_{b}},
    \end{equation}
    \begin{equation}\label{q7}
  \hat{g}_{q}=\sum_{M'_{a},M_{b},F_{b}}
  (g_{q})^{F'_{a}F_{b}}_{M'_{a}M_{b}}\cdot \hat{P}^{F'_{a}F_{b}}_{M'_{a}M_{b}},
    \end{equation}
    \begin{equation}\label{q8}
   (g_{cq})^{F_{a}F_{b}}_{M_{a}M_{b}} =
(-1)^{F_{a}-M_{a}}\left(\matrix{F_{a}&1&F_{b}
    \cr -M_{a}&q&M_{b}}\right) g_{F_{a}F_{b}},
    \end{equation}
    \begin{equation}\label{q9}
   (g_{q})^{F'_{a}F_{b}}_{M'_{a}M_{b}} =
(-1)^{F'_{a}-M'_{a}}\left(\matrix{F'_{a}&1&F_{b}
    \cr -M'_{a}&q&M_{b}}\right) g_{F'_{a}F_{b}},
    \end{equation}
    \begin{equation}\label{q10}
g_{F_{a}F_{b}} = (-1)^{F_{a}+J_{a}+I+1} G_{F_{a}F_{b}},
    \end{equation}
    \begin{equation}\label{q11}
G_{F_{a}F_{b}} = [(2F_{a}+1)(2F_{b}+1)]^{1/2}
\left\{\matrix{I&F_{a}&J_{a} \cr 1&J_{b}&F_{b}}\right\}.
    \end{equation}
Note, that the summation in (\ref{q6}) and (\ref{q7}) is carried out
over all possible values of the atomic total angular momentum
$F_{b}$ of the excited state.

After the action of the coherent pulse with the duration $T$ the
system density matrix, being initially $\hat{\rho}_{0}$, evolves to:
    \begin{equation}\label{q12}
   \hat{\rho}=\hat{S}\hat{\rho}_{0}\hat{S}^{+},
    \end{equation}
where the evolution operator $\hat{S}$ may be expressed through the
matrix exponent:
    \begin{equation}\label{q13}
   \hat{S}=\exp\left(\imath\hat{V}T\right)=
   \exp\left\{\imath\left(\hat{G}+\hat{G}^{+}\right)\right\},
    \end{equation}
    \begin{equation}\label{q14}
  \hat{G}=\theta_{c}\hat{p}_{c} - \imath \theta \hat{a}^{+}\hat{p},~
  \theta_{c}=\chi_{c}T,~\theta=\chi T.
    \end{equation}
With the use of the expansion of the exponent function in Taylor
series and with the use of relation
    \begin{equation}\label{q15}
(\hat{G}+\hat{G}^{+})^{2n}=\hat{Q}^{2n}+
\hat{G}\hat{Q}^{2(n-1)}\hat{G}^{+}~,~n=1,2,...~,
    \end{equation}
where
     \begin{equation}\label{q16}
\hat{Q}^{2}= \theta^{2}_{c}\hat{p}_{c}^{+}\hat{p}_{c}+
\theta^{2}\hat{a}\hat{a}^{+}\hat{p}^{+}\hat{p},
     \end{equation}
the evolution operator $\hat{S}$ may be transformed to the
expression:
     \begin{equation}\label{q17}
\hat{S}=\hat{P}_{F_{a}}+\hat{P}_{F'_{a}}+\hat{C}-\frac{1}{2} \hat{G}
\hat{F} \hat{G}^{+}+\imath \hat{G}\hat{H}+ \imath
\hat{H}\hat{G}^{+},
     \end{equation}
     \begin{equation}\label{q18}
\hat{C}=\cos(\hat{Q}),~ \hat{H}=\frac{\sin(\hat{Q})}{\hat{Q}},~
\hat{F}=\frac{\sin^{2}(\hat{Q}/2)}{(\hat{Q}/2)^{2}}.
     \end{equation}

\section{The probability of single-photon emission}

Initially the atom is at the equilibrium ground state, while the
cavity field is at the vacuum state, so that the initial system
density matrix is:
    \begin{equation}\label{q19}
\hat{\rho}_{0}=\frac{1}{2F_{a}+1}\hat{P}_{F_{a}}\cdot |0><0|,
    \end{equation}
$\hat{P}_{F_{a}}$ being the projector on the subspace of the
hyperfine-structure component $F_{a}$. The probability to detect a
single photon in the cavity is given by the formula:
     \begin{equation}\label{q20}
    w=Tr\{<1|\hat{\rho}|1>\},
     \end{equation}
where the trace is carried out in atomic variables. With the use of
the evolution operator (\ref{q17}) this probability may be written
as:
     \begin{equation}\label{q21}
    w=\frac{1}{(2F_{a}+1)}Tr\{\hat{R}\hat{R}^{+}\},
     \end{equation}
with
     \begin{equation}\label{q22}
\hat{R}=\frac{\theta\theta_{c}}{2}\hat{p}\frac{\sin^{2}
(\hat{Q}_{b}/2)}{(\hat{Q}_{b}/2)^{2}}\hat{p}_{c}^{+},
     \end{equation}
while the atomic operator
    \begin{equation}\label{q23}
\hat{Q}_{b}^{2}= \theta^{2}_{c}\hat{p}_{c}^{+}\hat{p}_{c}+
\theta^{2}\hat{p}^{+}\hat{p}
     \end{equation}
acts in the subspace of the excited level $b$. The operators
$\hat{R}$ in (\ref{q21}) may be also expressed in terms of atomic
operators
    \begin{equation}\label{q24}
\hat{Q}_{a}^{2}= \theta^{2}_{c}\hat{p}_{c}\hat{p}_{c}^{+} +
\theta^{2}\hat{p}\hat{p}^{+} + \theta \theta_{c}
(\hat{p}\hat{p}_{c}^{+} + \hat{p}_{c}\hat{p}^{+}),
    \end{equation}
acting in the subspace of the ground state levels $F_{a}$ and
$F'_{a}$:
     \begin{equation}\label{q25}
\hat{R}=\hat{P}_{F'_{a}} \cos(\hat{Q}_{a})\hat{P}_{F_{a}},
     \end{equation}
since
    \begin{equation}\label{q26}
\theta\theta_{c}\hat{p}\hat{Q}_{b}^{2n}\hat{p}_{c}^{+} =
\hat{P}_{F'_{a}}\hat{Q}_{a}^{2(n+1)}\hat{P}_{F_{a}},~n=0,1,2...
    \end{equation}
The presentation of operators $\hat{R}$ in (\ref{q21}) by means of
$\hat{Q}_{a}$ is more convenient here, than by means of
$\hat{Q}_{b}$, because the summation in (\ref{q24}) in all possible
values of the total angular momenta $F_{b}$ of the upper level and
its projections $M_{b}$ may be carried out analytically with the
help of the summation formulae for 3J- and 6J-symbols \cite{n10}.
After such summation the operator $\hat{Q}_{a}^{2}$ (\ref{q24})
becomes as follows:
    \begin{equation}\label{q27}
\hat{Q}_{a}^{2}= \theta^{2}_{c}\hat{A}_{c} + \theta^{2}\hat{A} +
\theta \theta_{c} (\hat{B} + \hat{B}^{+}),
    \end{equation}
    \begin{equation}\label{q28}
  \hat{A}_{c} = \sum_{M_{a},M'_{a}} (A_{c})^{F_{a}F_{a}}_{M_{a}M'_{a}} \cdot
  \hat{P}^{F_{a}F_{a}}_{M_{a}M'_{a}},
    \end{equation}
    \begin{equation}\label{q29}
  \hat{A} = \sum_{M_{a},M'_{a}} (A)^{F'_{a}F'_{a}}_{M_{a}M'_{a}} \cdot
  \hat{P}^{F'_{a}F'_{a}}_{M_{a}M'_{a}},
    \end{equation}
    \begin{equation}\label{q30}
  \hat{B} = \sum_{M_{a},M'_{a}} (B)^{F_{a}F'_{a}}_{M_{a}M'_{a}} \cdot
  \hat{P}^{F_{a}F'_{a}}_{M_{a}M'_{a}},~
    \end{equation}
    \begin{equation}\label{q31}
(B)^{F_{a}F'_{a}}_{M_{a}M'_{a}} = (-1)^{M'_{a}} \sum_{k,q}
\left(\matrix{ k & F_{a} & F'_{a} \cr q & M_{a} &
-M'_{a}}\right)a_{k}f^{k}_{q},
     \end{equation}
    \begin{equation}\label{q32}
a_{k}=(-1)^{F'_{a}-F_{a}+I-J_{b}}(2k+1)\left\{\matrix{ k & 1 & 1 \cr
J_{b} & J_{a} & J_{a}}\right\}b_{k},
     \end{equation}
     \begin{equation}\label{q33}
b_{k}=[(2F_{a}+1)(2F'_{a}+1)]^{1/2} \left\{\matrix{ k & F_{a} &
F'_{a} \cr I & J_{a} & J_{a}}\right\},
     \end{equation}
     \begin{equation}\label{q34}
f^{k}_{q}=\sum_{q_{1},q_{2}}(-1)^{q} (l_{c})_{-q_{1}}(l)_{-q_{2}}
\left(\matrix{ k & 1 & 1 \cr q & q_{1} & q_{2}}\right).
     \end{equation}
The matrix elements $(A)^{F'_{a}F'_{a}}_{M_{a}M'_{a}}$ are obtained
from $(B)^{F_{a}F'_{a}}_{M_{a}M'_{a}}$ by the substitutions $A
\rightarrow B$, $F'_{a} \rightarrow F_{a}$ elsewhere in
(\ref{q31})-(\ref{q33}) and $l\rightarrow l_{c}$ in (\ref{q34}),
while elements $(A_{c})^{F_{a}F_{a}}_{M_{a}M'_{a}}$ are obtained
from $(B)^{F_{a}F'_{a}}_{M_{a}M'_{a}}$ by the substitutions $A_{c}
\rightarrow B$, $F_{a} \rightarrow F'_{a}$ in
(\ref{q31})-(\ref{q33}) and $l_{c}\rightarrow l$ in (\ref{q34}). Now
the probability (\ref{q21}) may be calculated by reducing the
hermitian $2(F_{a}+F'_{a}+1)\times 2(F_{a}+F'_{a}+1)$ matrix $Q_{a}$
to its diagonal form.

\begin{figure}[t]
\includegraphics[width=7cm]{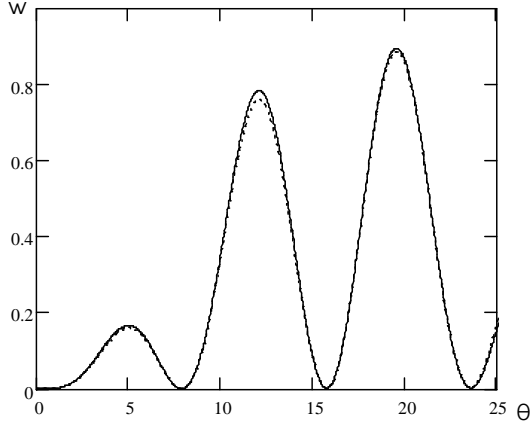}
\caption{Probability $w$ of single photon emission versus reduced
Rabi angle $\theta$ = $\theta_{c}$ at the orthogonal polarizations
$\psi=90^{\circ}$ of the cavity and laser fields. Solid line refers
to the angular momenta $F_{a}=2$, $F'_{a}=3$, $I=5/2$, $J_{a}=1/2$,
$J_{b}=3/2$, while the dashed line refers to $F_{a}=3$, $F'_{a}=4$,
$I=7/2$, $J_{a}=1/2$, $J_{b}=3/2$.}
\end{figure}

The matrices $\hat{A}$, $\hat{A}_{c}$ and $\hat{B}$ in (\ref{q27})
simplify essentially in case, when the electronic angular momentum
of the atomic ground state equals $1/2$: $J_{a}=1/2$, which is the
usual case for alkali metals. In this case we obtain:
$$\hat{A}_{c}=\frac{1}{6} \hat{P}_{F_{a}},~
\hat{A}=\frac{1}{6}\hat{P}_{F'_{a}},$$
$$(B)^{F_{a}F'_{a}}_{M_{a}M'_{a}}=\imath \sin(\psi)
\delta_{M_{a},M'_{a}} B(F_{a},F'_{a},M_{a}),$$ where $\psi$ is the
angle between polarization vectors of the coherent pulse ${\bf
l_{c}}$ and of the cavity field ${\bf l}$, while functions
$B(F_{a},F'_{a},M_{a})$ are defined from (\ref{q31})-(\ref{q34}).
So, in this case the photon will not be emitted in the cavity $w=0$,
if the polarizations of the coherent pulse and of the photon are
collinear: $\psi=0$. The dependence of the probability $w$ of single
photon emission on the reduced vacuum Rabi angle $\theta$ of the
cavity field, which is considered to be equal to the reduced area
$\theta_{c}$ of the coherent laser pulse ($\theta$=$\theta_{c}$), at
the orthogonal polarizations $\psi=90^{\circ}$ of the cavity and
laser fields is presented at Fig.2. The solid line refers to the
values of the angular momenta $F_{a}=2$, $F'_{a}=3$, $I=5/2$,
$J_{a}=1/2$, $J_{b}=3/2$ corresponding to the transitions in
$^{85}Rb$, while the dashed line refers to $F_{a}=3$, $F'_{a}=4$,
$I=7/2$, $J_{a}=1/2$, $J_{b}=3/2$ corresponding to the transitions
in $^{133}Cs$. In both cases the maximum probability values are
obtained at the same value of Rabi angles $\theta_{max}=19.604$,
though the values of probability at $\theta=\theta_{max}$ slightly
differ: $w_{max}=0.8943$ for Rubidium transitions and
$w_{max}=0.8855$ for Cesium transitions. The greater probabilities
are obtained, if the coherent laser pulse in course of photon
emission drives the atom from the less degenerate hyperfine
structure component $F_{a}$ to the more degenerate one
$F'_{a}>F_{a}$, otherwise the probabilities are decreased by the
factor $(2F_{a}+1)/(2F'_{a}+1)$. However the maximum emission
probabilities occur not at the orthogonal polarizations of laser and
cavity fields. The Fig.3 represents the dependence of the
probability $w$ of single photon emission at Rabi angle
$\theta$=$\theta_{c}$=$\theta_{max}=19.604$, corresponding to the
maximum probability values, on the angle $\psi$  between
polarization vectors of the laser pulse and of the cavity field. The
solid line, like at Fig.2, refers to Rubidium transitions with
$F_{a}=2$, $F'_{a}=3$, $I=5/2$, $J_{a}=1/2$, $J_{b}=3/2$, while the
dashed line refers to Cesium transitions with $F_{a}=3$, $F'_{a}=4$,
$I=7/2$, $J_{a}=1/2$, $J_{b}=3/2$. As it may be seen from Fig.3 the
maximum probabilities of the photon emission are obtained at the
angle $\psi=60^{\circ}$ and the maximum probability values are:
$w_{max}=0.9595$ for Rubidium transitions and $w_{max}=0.9397$ for
Cesium transitions.

\begin{figure}[t]
\includegraphics[width=7cm]{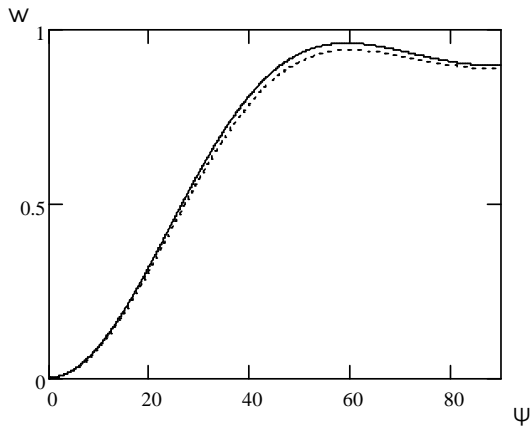}
\caption{Probability $w$ of single photon emission versus angle
$\psi$ between polarization vectors of the laser pulse and of the
cavity field at the reduced Rabi angle
$\theta$=$\theta_{c}$=$\theta_{max}=19.604$. Solid line refers to
the angular momenta $F_{a}=2$, $F'_{a}=3$, $I=5/2$, $J_{a}=1/2$,
$J_{b}=3/2$, while the dashed line refers to $F_{a}=3$, $F'_{a}=4$,
$I=7/2$, $J_{a}=1/2$, $J_{b}=3/2$.}
\end{figure}

\section{Conclusions}

In the present paper we derived the expression for the probability
of single photon emission via Raman scattering of short laser pulse
under the conditions when the hyperfine structure of the atomic
ground state is totally resolved, while the the hyperfine structure
of the excited state is totally unresolved, for arbitrary values of
the level angular momenta. In case of the ground state electronic
angular momentum $J_{a}=1/2$, which is the usual case for the
experiments on deterministic single photon emission in alkali
metals, the emission probability strongly depends on the angle
between polarizations of the laser pulse and of the quantized cavity
field - this probability is zero at the collinear polarizations of
the cavity and laser fields and it obtains its maximum values close
to unity, when the angle between their polarizations constitutes
$60^{\circ}$. The greater emission probabilities are obtained, if
the coherent laser pulse drives the atom from the less degenerate
hyperfine structure component to the more degenerate one in course
of photon emission.

{\bf Acknowledgements}

Authors are indebted for financial support of this work to Russian
Foundation for Basic Research (grant 08-02-00161).

\end{document}